\documentclass[aps,showpacs,amsmath,amssymb,12pt]{revtex4}
\usepackage{latexsym}
\usepackage{bm}

\def\HollowBox #1#2{{\dimen0=#1 \advance\dimen0 by -#2
       \dimen1=#1 \advance\dimen1 by #2
        \vrule height #1 depth #2 width #2
        \vrule height 0pt depth #2 width #1
        \llap{\vrule height #1 depth -\dimen0 width \dimen1} 
       \hskip -#2
       \vrule height #1 depth #2 width #2}}
\def\BOX{\HollowBox{.100in}{.010in}}

\begin{document}

\title{Minimal  Massive Gravity: Conserved Charges, Excitations and the Chiral Gravity Limit }

\author{Bayram Tekin}  
\email{btekin@metu.edu.tr}
\affiliation{Department of Physics,\\
             Middle East Technical University, 06800, Ankara, Turkey}

\date{\today}

\begin{abstract}
We find the excitations and construct the  conserved charges ( mass and angular momentum) of the recently found Minimal Massive Gravity (MMG) in $2+1$ dimensions in asymptotically Anti-de Sitter (AdS) spacetimes. The field equation of the theory does not come from an action and hence lacks the required Bianchi Identity needed to define conserved charges. But the theory, which also provides a healthy extension of the Topologically Massive Gravity in the bulk and boundary of spacetime, does admit conserved charges for metric that are solutions.  Our construction is based on background Killing vectors and imperative to provide physical meaning to the integration constants in the black hole type metrics.  We also study the chiral gravity limit of MMG. 

\end{abstract}

\maketitle

\section{\label{intro} Introduction}

It is well-known that Einstein's gravity in  $2+1$  dimensions is devoid of any local degrees of freedom even though it has black hole solutions when a negative cosmological constant is added \cite{btz}.[We shall work with the mostly plus signature.]  On the other hand the parity-violating Topologically Massive Gravity (TMG)  \cite{djt}  has both a single massive spin-2  local degree of freedom as well as various black hole solutions.  But TMG has a bulk-boundary unitarity conflict which makes it rather unsuitable for the AdS/CFT "applications". Namely,  either the bulk or the 
boundary theory is non-unitary as can be seen from the comparison of the unitarity regions defined by the positivity of the two Brown-Henneaux boundary c-charges \cite{bh} and the bulk energies \cite{dt_tmg}. In a certain parameter choice, one hoped that one has a unitary "chiral gravity" theory \cite{strom} but it turns out that at the chiral point there appear Log solutions making the boundary theory a non-unitary  logarithmic CFT \cite{carlip,grumiller,gribet}. 
Unfortunately, this state of affairs, (the bulk-boundary unitarity conflict) remained intact in the "New Massive Gravity" (NMG) \cite{nmg}  that  provided a non-linear extension of the three-dimensional Fierz-Pauli theory with two spin-2 degrees of freedom both in flat and AdS spacetimes \cite{canon}.  Further modification of NMG with more powers of curvature did not solve the unitary conflict \cite{sinha, binmg,paulos}. For example, Born-Infeld extension of NMG with (in principle) infinitely many powers of curvature with rather remarkable properties is unitary either in the bulk or on the boundary of AdS \cite{binmg, binmgc}.  

At this point one must be quite puzzled: Einstein's gravity is healthy both in the bulk and boundary but it has no local degrees of freedom and hence in some sense not a good example where one can study "quantum gravity". On the other hand, the above-mentioned many, otherwise healthy,  non-trivial theories with one or two massive gravitons, fail to be   unitary on the boundary. Apparently, one is forced to choose either local-triviality or boundary non-unitarity. In the age of AdS/CFT, the second option is nothing less than heresy,  hence we are back with Einstein's theory. But, recently, a very interesting paper \cite{townsend} appeared  in which the authors have shown that there is one more virtue, slightly less dangerous than the previous two, that one can let go: that is the Lagrangian formulation of the theory. Namely, they defined a theory-Minimal Massive Gravity (MMG)- which does not come from the variation of an action with the metric as the independent field and hence 
 also lacks the all important Bianchi Identity and the diffeomorphism invariance.  But nevertheless, a consistent restricted version of the theory has a single massive degree of freedom that is unitary in the bulk and gives rise to a unitary CFT on the boundary. 
[See also the recent work on MMG which couples MMG to matter fields ( a non-trivial construction ) \cite{townsend2} and \cite{baykal} for another derivation of the 
theory.] All this is quite good since after all MMG seems to be the theory without the infamous bulk-boundary unitarity conflict. But of course, to make sense of the black hole solutions of the theory, to define thermodynamics { \it{etc} } one has to have conserved charges, especially energy in the theory. Otherwise, one could not even assign a physical meaning to the parameters ( or integration constants ) in the solutions of the theory. 

This is the task we carry out in this work following the well-established ADM \cite{adm} charges of General Relativity (or in retrospect any purely metric-based gravity theory for asymptotically flat spacetimes)  and  AD \cite{ad} charges of (cosmological ) General Relativity  for asymptotically (A)dS spacetimes  and their generalization, the ADT \cite{adt} charges, for quadratic gravity theories and generic $ f(\mbox{Riemann})$ theories \cite{amsel,senturk}. The details of this Killing charge construction is given elsewhere  \cite{adt} hence the discussion in section II. will be just a recapitulation of the essential points.

We also study the excitations of the MMG theory in AdS backgrounds, construct the boundary central charges, and excitation energies as well as the chiral gravity limit of the theory.

\section{\label{conserved} The conserved charges of gravity theories endowed with Bianchi Identities }

Given the field equations of the theory as 

\begin{equation}
\Phi_{\mu\nu}(g,R,\nabla \mbox{Riemann},R^{2},...)=\kappa\tau_{\mu\nu},\label{gen_denk}
\end{equation}
one assumes that  for vanishing $\tau_{\mu \nu}$, (A)dS is  the background  solution, $\Phi_{\mu\nu}(\bar{g},\bar{R},\bar{\nabla}\bar{R},\bar{R}^{2}...)=0$, with the metric $\bar{g}_{\mu \nu}$ with vanishing energy and angular momenta (let us stay in generic $n$-dimensions for this brief discussion, so there could be more than one angular momentum).  (A)dS has the maximum number of symmetries, let us denote the Killing vectors as $\bar{\xi}_\mu$. ( Note that not to clutter the notation, we do not put another index on the Killing vector that could identify the different Killing vectors.) 
Given a spacetime with the metric $g_{\mu \nu}$,  that asymptotically has the same Killing symmetries as the background space, we can define "conserved 
charges"  by first linearizing the field equations as

\begin{equation}
{\cal O}(\bar{g})_{\mu\nu\alpha\beta}h^{\alpha\beta}=\kappa \tau_{\mu\nu}, \label{lin_denk}
\end{equation}
with  $g_{\mu\nu}\equiv \bar{g}_{\mu\nu}+ h_{\mu\nu}$ and the operator ${\cal O}(\bar{g})_{\mu\nu\alpha\beta}$ , a $(0,4)$ tensor in AdS can be easily found given the field equations.   Note that the right hand-side of  (\ref{lin_denk}) has all the terms in the metric perturbation ($h_{\mu \nu}$ ) save the linear one. If the full equation (\ref{gen_denk}) comes from a diffeomorphism-invariant action then it satisfies the full Bianchi Identity  $\nabla_\mu \Phi^{\mu\nu}=0$  with the covariant derivative taken with respect to the metric compatible $g_{\mu \nu}$. This then leads to the background "Bianchi Identity" or background (covariant ) conservation of the linearized equation $\bar{\nabla}_\mu {\cal O}(\bar{g})^\mu\,_{\nu\alpha\beta}h^{\alpha\beta}=0$.
This is not yet sufficient to define globally conserved charges, one makes use of the Killing vectors to define a partially conserved current via 
$ \sqrt{-\bar{g}} \bar{\nabla}_\mu ( \bar{\xi}_\nu T^{\mu \nu})  =    \partial_\mu(\sqrt{-\bar{g}} \bar{\xi}_\nu T^{\mu \nu} )=0 $.  Using the Stokes' theorem one arrives at the following conserved global charges for each Killing vector 
\begin{equation}
Q^\mu\left ( \bar{\xi} \right) =\int_{\cal M }d^{n-1} x\sqrt{-\bar{g}}\bar{\xi}_\nu T^{\mu \nu} = \int_{ \Sigma }d \Sigma_i {\cal F}^{\mu i}
\end{equation}
where ${\cal M }$ is our $n-1$ dimensional spatial manifold with  $\Sigma$ being its boundary.  Here, a crucial step for each theory is to find the anti-symmetric tensor ${\cal F}^{\mu\nu}$ that satisfies  $T^{\mu\nu}\bar{\xi}_{\nu}=\bar{\nabla}_{\nu}{\cal F}^{\mu\nu}$.

This procedure has  been successfully applied to many theories where the assumed conditions on asymptotic symmetries and the implicit assumption of the proper fall of the perturbation $h_{\mu \nu}$ at the boundary are satisfied.  For example, let us recall the conserved charges of TMG that is pertinent to our main discussion of MMG charges. The TMG field equations read 
\begin{equation}
 R_{\mu\nu} - \frac{1}{2} \, g_{\mu\nu} \, R + \Lambda \, g_{\mu\nu} 
+ \frac{1}{\mu} \, C_{\mu\nu} = 0 \, 
\end{equation}
with the Cotton tensor given as
\begin{equation}
C^{\mu\nu} \equiv \frac{1}{\sqrt{-g}} \, \epsilon^{\mu\alpha\beta} \, 
\nabla_{\alpha} S^\nu_ \beta\, , \label{tmg} 
\end{equation}
where $S^\nu_\beta= R^{\nu}_{\beta} - \frac{1}{4}
\delta^{\nu}_{\beta} \, R $ is the Schouten tensor in 3 dimensions.  The field equations of TMG come from an action which is diffeomorphism-invariant up to a boundary term and hence the theory is endowed with the Bianchi Identity and amenable to our charge definition. This was done in \cite{dt_tmg} with the following result 

\begin{equation}
Q^{\mu} (\bar{\xi}) = \frac{1}{2 \pi G_3} \, \oint_{\partial {\Sigma}} \,
dl_{i} \, \left( q^{\mu i}_{E} (\bar{\xi}) + \frac{1}{2 \mu} \, 
q^{\mu i}_{E} (\bar{\Xi}) + \frac{1}{2 \mu} \, q^{\mu i}_{C} (\bar{\xi})
\right) \, , \label{charge}
\end{equation}
where the parts coming from the Einstein tensor and the Cotton tensor read respectively as 
\begin{eqnarray}
q^{\mu i}_{E} (\bar{\xi}) & \equiv & \sqrt{-\bar{g}} \left( 
\bar{\xi}_{\nu} \, \bar{\nabla}^{\mu} \, h^{i \nu} -
\bar{\xi}_{\nu} \, \bar{\nabla}^{i} \, h^{\mu\nu} +
\bar{\xi}^{\mu} \, \bar{\nabla}^{i} \, h -
\bar{\xi}^{i} \, \bar{\nabla}^{\mu} \, h \right.  \nonumber \\
& & \quad \qquad \left. + h^{\mu\nu} \, \bar{\nabla}^{i} \, \bar{\xi}_{\nu}
- h^{i \nu} \, \bar{\nabla}^{\mu} \, \bar{\xi}_{\nu}
+ \bar{\xi}^{i} \, \bar{\nabla}_{\nu} \, h^{\mu\nu}
- \bar{\xi}^{\mu} \, \bar{\nabla}_{\nu} \, h^{i \nu}
+ h \, \bar{\nabla}^{\mu} \, \bar{\xi}^{i} \right) \, , \label{einsteincharge} \\
q^{\mu i}_{C} (\bar{\xi}) & \equiv & 
\epsilon^{\mu i \beta} \, {\cal G}_{\nu\beta} \, \bar{\xi}^{\nu}
+ \epsilon^{\nu i \beta} \, {\cal G}^{\mu}\,_{\beta} \, \bar{\xi}_{\nu}
+ \epsilon^{\mu\nu\beta} \, {\cal G}^{i}\,_{\beta} \, \bar{\xi}_{\nu}.
\label{cottoncharge}
\end{eqnarray}
Here, interestingly, a new Killing vector built  out of the curl of the  background Killing vector arises  : \( \bar{\Xi}^{\beta} \equiv \epsilon^{\alpha\nu\beta} \, 
\bar{\nabla}_{\alpha} \, \bar{\xi}_{\nu} / \sqrt{-\bar{g}} \). All contractions and raising and lowering must be done with the background metric, for example
$ h= \bar{g}^{\mu \nu}h_{\mu \nu}$. Even though, the background tensors that appeared here were defined in \cite{dt_tmg,olmez}, it pays to collect them here as we shall need some of them below for the computation in the  MMG  theory. 
The background satisfies
\begin{equation}
 \bar{R}_{\mu\alpha\nu\beta} = \Lambda \, 
(\bar{g}_{\mu\nu} \, \bar{g}_{\alpha\beta} - 
\bar{g}_{\mu\beta} \, \bar{g}_{\alpha\nu}) \, , \quad
\bar{R}_{\mu\nu} = 2 \Lambda \, \bar{g}_{\mu\nu} \, , \quad
\bar{R} = 6 \Lambda 
\end{equation}
 In 3 dimensions, we do not need the linearization of the Riemann tensor, hence, respectively, linearized Ricci tensor, and the Ricci scalar read
\[ R_{\mu\nu}^{L} = \frac{1}{2} (- \bar{\BOX} \, {h}_{\mu\nu} 
- \bar{\nabla}_{\mu} \, \bar{\nabla}_{\nu} \, h + \bar{\nabla}^{\sigma} \,
\bar{\nabla}_{\nu} \, h_{\sigma\mu} + \bar{\nabla}^{\sigma} \,
\bar{\nabla}_{\mu} \, h_{\sigma\nu}) \, , \]
\[ R^{L} \equiv (R_{\mu\nu} \, g^{\mu\nu})^{L} = 
R_{\mu\nu}^{L} \, \bar{g}^{\mu\nu} - 2 \Lambda \, h = - \bar{\BOX} \, h
+ \bar{\nabla}_{\mu} \, \bar{\nabla}_{\nu} \, \bar{h}^{\mu\nu} 
- 2 \Lambda \, h \, , \]
These can be  used to find the linearized cosmological Einstein and the
Cotton tensors as
\begin{eqnarray*}
{\cal G}_{\mu\nu} & \equiv & (G_{\mu\nu} +\Lambda g_{\mu \nu})^{L} =
R_{\mu\nu}^{L} - \frac{1}{2} \, \bar{g}_{\mu\nu} \,
R^{L} - 2 \Lambda \, {h}_{\mu\nu} \, , \\
C^{\mu\nu}_{L} & = & \frac{1}{\sqrt{-\bar{g}}} \, \epsilon^{\mu\alpha\beta} \,
\bar{g}_{\beta\sigma} \, \bar{\nabla}_{\alpha} \, \left( R^{\sigma\nu}_{L}
- 2 \, \Lambda \, h^{\sigma\nu} - \frac{1}{4} \, \bar{g}^{\sigma\nu} 
\, R_{L} \right) \, . 
\end{eqnarray*}
Here, as usual,  $G_{\mu\nu} \equiv R_{ \mu \nu}  -\frac{1}{2}g_{\mu \nu} R $.

Before we conclude this section and move on to our main goal, let us note that for a time like Killing vector $\bar{\xi}^\mu = (-1,0,0)$ , $Q^0$ corresponds to the energy which is background diffeomorphism invariant only if the spatial boundary is at infinity ( as in the case of flat and (A)dS space ).  For ${\bar \xi}^\mu = (0,0,1)$  (say in polar coordinates) vectors, $Q^0$ is the angular momentum. (Please note that the construction is coordinate independent ).  See some example computations in \cite{kanik,olmez}.

\section{\label{mmg}  Excitations, Conserved Charges and Chiral Gravity limit of MMG   }

\subsection{\label{lin} {Linearization of the Field equations: Excitations}}
\
As explained in the Introduction, MMG theory was designed to be free of the the bulk-boundary unitarity conflict. But to obtain unitarity everywhere, the authors of \cite{townsend} bartered unitarity with the precious Lagrangian formulation  and hence the Bianchi Identity that must be valid for any metric is gone.  This of course is quite worrisome in terms of the conserved charge definition as we stressed above. But there is a resolution as we shall see.
The field equations of MMG are 
\begin{equation}
E_{\mu\nu} \equiv \Lambda_0\,  g_{\mu\nu}  + \sigma G_{\mu\nu}  + \frac{1}{\mu} C_{\mu\nu} + \frac{\gamma}{\mu^2} J_{\mu\nu} = 0\, , 
\label{mmg_denk}
\end{equation}
with two dimensionless parameters $\sigma$ and $\gamma$ as well as two dimensionful ones $\mu$ and $\Lambda_0$.  
The new ingredient is the $J$ tensor defined  as  
\begin{equation}
J^{\mu\nu} \equiv \frac{1}{2\det g} \, \varepsilon^{\mu\rho\sigma}\varepsilon^{\nu\tau\eta} S_{\rho\tau}S_{\sigma\eta}\, ,
\end{equation}
one has a non-vanishing covariant divergence for generic metrics :
\begin{equation}
\sqrt{-\det g}\, \nabla_\mu J^{\mu\nu} =    \varepsilon^{\nu\rho\sigma} S_\rho{}^\tau C_{\sigma\tau}\, .
\end{equation}
This is at the root of the problem, but as noted in \cite{townsend,townsend2}, for the {\it solutions } of the theory, this is indeed zero. Therefore  we can define conserved charges.  First let us note that one can rewrite the $J$-tensor as
\begin{equation}
J_{\mu \nu} = G_\mu^\rho G_{\rho \nu} - \frac{1}{2} g_{\mu \nu} G_{\rho \sigma} G^{\rho \sigma} + \frac{1}{4}G_{\mu \nu} R + \frac{1}{16} g_{\mu \nu} R^2. \label{J-ten}
\end{equation}
We can now find the effective cosmological constant of the theory (\ref{mmg_denk}) by first noting that $\bar{J}^{\mu \nu}=\frac{\Lambda^2}{4} \bar{g}^{\mu \nu} $. So the vacuum field equation reads
\begin{equation}
\Lambda_0 - \sigma \Lambda + \frac{\gamma}{4 \mu^2} \Lambda^2 =0,
\label{vac}
\end{equation}
with solutions
\begin{equation}
\Lambda_\pm = \frac{2 \mu^2}{\gamma} \left( \sigma \pm \sqrt{ \sigma^2 - \frac{\gamma \Lambda_0}{\mu^2}} \, \right)
\end{equation}
We agreed that for the solutions of the full theory we have 
\begin{equation}
 \nabla_\mu J^{\mu\nu} =0,
\end{equation}
whose linearization about the (A)dS vacuum leads to a background conserved tensor
\begin{equation}
\bar{\nabla}_\mu {\cal{J}}_L^\mu =0.
\end{equation}
where  ${\cal{J}}_L^{\mu \nu} \equiv  (J^{\mu \nu} )_L + \frac{\Lambda^2}{4} h^{\mu \nu} $. Note the all important second term which makes the total expression a background diffeomorphism invariant expression under  transformations $ \delta_\xi h_{\mu \nu} = \bar{\nabla}_\mu \zeta_\nu  + \bar{\nabla}_\nu \zeta_\mu$. 
Let us now compute the linearized form of  (\ref{J-ten}) with the help of the linearized tensors given above to get
\begin{equation}
(J^{\mu \nu})_L = - \frac{\Lambda}{2} {\cal G}^{\mu\nu}  - \frac{\Lambda^2}{4} h^{\mu \nu},
\end{equation}
hence one has  ${\cal J}^{\mu \nu}_L = - \frac{\Lambda}{2} {\cal G}^{\mu\nu}$ which is needed in defining the conserved charges of the full theory (\ref{mmg_denk}) whose linearization about one of its (A)dS vacua gives
\begin{equation}
\bar{E}_{\mu \nu} + (\Lambda_0 - \sigma \Lambda ) h_{\mu \nu} +  \sigma  {\cal G}_{\mu\nu} + \frac{1}{\mu} C^L_{\mu \nu}
+ \frac{\gamma}{\mu^2} (J_{\mu \nu})^L  \equiv \kappa T_{\mu \nu},
\end{equation}
where the right hand-side represents all the non-linear terms in $h$. We have also introduced a scaled Newton's constant ($\kappa$) to keep the conventional dimensions of the  energy-momentum tensor. Since $ (J_{\mu \nu})^L = - \frac{\Lambda}{2} {\cal G}_{\mu\nu}  + \frac{\Lambda^2}{4} h^{\mu \nu}$, using the vacuum field equation in the first term and the terms multiplying $h_{\mu \nu}$, we arrive at
\begin{equation}
\left (\sigma -  \frac{\gamma \Lambda }{2 \mu^2} \right)  {\cal G}_{\mu\nu} + \frac{1}{\mu} C^L_{\mu \nu}
 = \kappa T_{\mu \nu},
\label{lin_ark}
\end{equation}
which is nothing but the linearized field equations of TMG with a modified coefficient in front of the Einsteinian part.   Before we turn back to the conserved charge issue, let us say a few words about the bulk excitations of the theory: The linearized equation about (A)dS explains why bulk properties of MMG are the same as TMG, albeit with a modified mass: From  (\ref{lin_ark}), using the results of the previous works \cite{carlip, gursess}, we can write the mass of the single spin 2-mode in MMG as
\begin{equation}
M_{\mbox{g}}^2=   \mu^2 \left (\sigma -  \frac{\gamma \Lambda }{2 \mu^2} \right)^2 + \Lambda ,
\end{equation}
which satisfies the Breitenlohner-Freedman (BF) bound \cite{bf}  $M_{\mbox{g}}^2 \ge \Lambda$ in AdS.  The graviton mass vanishes at the  two "chiral points" (a opposed to the one in TMG)  for AdS
\begin{equation}
\Lambda_0^\pm =\frac{\mu^2}{\gamma^3}\left (\gamma  \sigma (2 + \gamma \sigma) - 2  \pm 2 \sqrt{1 - 2 \gamma \sigma} \right).
\end{equation}
In fact, to actually see the "chirality" of the boundary conformal field theory, let us compute the left and right central charges in the theory. This was done in 
\cite{townsend}, but our notation and conventions are quite different and hence it actually pays to repeat the calculation here  since it is quite simple using our formulation. We already know the central charges of TMG and it is clear that as far as the central charges of the boundary theory is concerned,linearized MMG (\ref{lin_ark}) with a vanishing left hand-side is sufficient. Hence the two copies of the Virasoro algebra on the boundary has the following central charges
\begin{equation}
c_L = \frac{ 3 \ell}{2 G_3} \left ( \sigma + \frac{ \gamma}{2 \mu^2 \ell^2 } - \frac{1}{\mu \ell} \right ),  \hskip 1 cm 
 c_R = \frac{ 3 \ell}{2 G_3} \left ( \sigma + \frac{ \gamma}{2 \mu^2 \ell^2 } + \frac{1}{\mu \ell} \right ), \label{central}
\end{equation}
where we have used the AdS radius defined as $\ell^2 = -\frac{1}{\Lambda}$ and normalized the central charges as Brown and Henneaux \cite{bh}.

Let us now consider the energies of the linear excitations following \cite{strom,tahsin}.  For this purpose we need to find the (free) action leading to (\ref{lin_ark}) which is
\begin{equation}
S =\frac{1}{2 \pi G_3} \int d^3  x \sqrt{-\bar{g}} \left [ -\frac{1}{2} \left(\sigma -  \frac{\gamma \Lambda }{2 \mu^2} \right) h^{\mu \nu} {\cal G}_{\mu\nu} - \frac{1}{2 \mu} h^{\mu \nu} C^L_{\mu \nu} \right ].
\end{equation}
We need to get the Ostrogradsky Hamiltonian, before that it is a good idea to fix the gauge by choosing  the transverse traceless conditions 
\begin{equation}
\bar{\nabla}_\mu h^{\mu \nu} = 0,  \hskip 1 cm h=0,
\end{equation}
which reduces the action to
\begin{equation}
S =  -\frac{1}{4 \pi G_3} \int d^3  x \sqrt{-\bar{g}} \left [  \left(\sigma -  \frac{\gamma \Lambda }{2 \mu^2} \right ) \nabla_\alpha h_{\mu \nu} \bar{ \nabla}^\alpha h^{\mu \nu} + 2 \Lambda h_{\mu \nu} h^{\mu \nu} + \frac{1}{\mu} \epsilon_{\mu}^{\alpha \beta} \nabla_\alpha h_{\mu \nu} ( \bar{\Box} - 2 \Lambda ) h_{\beta \nu} 
\right ].
\end{equation} 
Choosing the AdS metric as
\begin{equation}
ds^2 = \frac{1}{- \Lambda}  \left ( -\cosh^{2}\rho dt^{2}+d\rho^{2}+\sinh^{2}\rho d\phi^{2} \right ), 
\end{equation}
and decomposing the metric into massive, left moving and right moving fluctuations as was done in \cite{strom}
\begin{equation}
h_{ \mu \nu} \equiv h_{\mu \nu}^M +  h_{\mu \nu}^L + h_{\mu \nu}^R,
\end{equation}
the Ostrogradsky Hamiltonian leads to following excitation energies 

\begin{equation}
E_M = \frac{ M_{\mbox{g}}^2}{4 \pi G_3} \frac{1}{T} \int d^3 x \, \sqrt{-\bar{g}} \,\epsilon_\alpha\,^{0\mu} h^{\alpha \nu}\partial_t h^M_{\mu \nu}, \label{energy1}
\end{equation}
\begin{equation}
E_L= -\frac{ c_L} {6 \pi \ell } \frac{1}{T} \int d^3 x \, \sqrt{-\bar{g}} \,\bar{\nabla}^0  h^{\alpha \nu}_L \partial_t h^L_{\mu \nu},
\label{en2}
\end{equation}
\begin{equation}
E_R= -\frac{ c_R} {6 \pi \ell } \frac{1}{T} \int d^3 x \, \sqrt{-\bar{g}} \,\bar{\nabla}^0  h^{\alpha \nu}_R \partial_t h^R_{\mu \nu}
\label{en3}
\end{equation}
where we have made use of the central charges (\ref{central}) and defined a (large) time  $T$ whose relevance is explained in \cite{tahsin}.  
To judge the positivity or the negativity of all these three energies we must find all solutions of the linearized theory which split into 3 equations: One for the massive 
\begin{equation}
\epsilon_\mu\,^{\alpha \beta} \bar{\nabla}_\alpha h^M_{\beta \nu} + \mu  \left(\sigma -  \frac{\gamma \Lambda }{2 \mu^2} \right ) h^M_{\mu \nu}=0,
\end{equation} 
and the other two for left and right moving modes
\begin{equation}
\epsilon_\mu\,^{\alpha \beta} \bar{\nabla}_\alpha h^L_{\beta \nu} +  \ell  h^L_{\mu \nu}=0, \hskip 1 cm  \epsilon_\mu\,^{\alpha \beta} \bar{\nabla}_\alpha h^R_{\beta \nu} -  \ell  h^R_{\mu \nu}=0,
\end{equation}
Fortunately,  the solutions of these equations were given in \cite{strom} using the $ SL(2,R)_L \times  SL(2,R)_R $  isometry of the AdS metric. Without going into further details, let us note that all the solutions furnish a representation of this group and can be generated from the  3 primaries with weights  $(h, \bar{h})$ . The left moving mode has  weights $(2,0)$ and the right moving mode has weights $(0,2)$ and the massive mode has  weights 
\begin{equation}
 h = \frac{3}{2} + \frac{\sigma \ell}{2} +  \frac{\gamma}{4  \ell \mu^2}, \hskip 1 cm  \bar{h} = -\frac{1}{2} + \frac{\sigma \ell}{2} +  \frac{\gamma}{4  \ell \mu^2},
\end{equation}
where we assumed  $\sigma -  \frac{\gamma \Lambda }{2 \mu^2}> 0$. We do not depict the explicit solution given in \cite{strom}, since we only the need the all important conclusion that once the solutions are plugged in to the energy expressions (\ref{energy1})to (\ref{en3}) all the integrals yield negative values.  This is quite good news for the left and right moving modes since than, as noted in \cite{townsend} both central charges can be positive and positivity of central charges is not in conflict with the excitation energies of the left and right moving modes. For the massive mode, assuming $\mu >0$, positivity of the energy demands that the square of the graviton mass is negative. But this is allowed in AdS, and does not lead to tachyons as long as the BF condition is satisfied. In his case as we have shown above, it is indeed satisfied. So we have a stronger condition than the Bf condition for positive energy massive modes
\begin{equation}
 0 \ge M^2 \ge  \Lambda.
\end{equation}
Therefore, unlike case of the TMG, there is no conflict between the bulk and boundary unitarity in MMG as noted in \cite{townsend}.
let us now look at the chiral point where $M^2_{\mbox{g}}=0$ hence the bulk graviton disappears, as well as the left moving modes with $c_L=0$. Right moving modes survive with positive energy and a central charge $c_R = \frac{3}{G_3 \mu}$ which differs from that of   the chiral gravity limit of TMG. 
One might worry about the existence of the parameter region which could allow  a positive $\mu$ under these conditions. It is easy to see that for $\gamma > 0$, the region $\sigma \gamma \le 1/2 $ yields such a $\mu$. There is another problem which we do not deal here that is whether the chiral gravity point is "really" unitary in the sense that there will be logarithmic solutions that will lead to a non-unitary logarithmic field theory as in the case of  chiral gravity limit of TMG \cite{grumiller}.  For New Massive Gravity see the analogous discussions in \cite{yan}

\subsection{\label{lin} Conserved Charges }

Having found the mass of the bulk spin-2 excitation, let us now return to our construction of conserved charges. With the background knowledge given in the previous section,we can now write down the conserved charges for the MMG theory
\begin{equation}
Q^{\mu} (\bar{\xi}) = \frac{1}{2 \pi G_3} \, \oint_{{\Sigma}} \, dl_{i}  \left(  (\sigma -  \frac{\gamma \Lambda }{2 \mu^2} )  q^{\mu i}_{E} (\bar{\xi}) + \frac{1}{2 \mu} \, 
q^{\mu i}_{E} (\bar{\Xi}) + \frac{1}{2 \mu} \, q^{\mu i}_{C} (\bar{\xi})
\right) \, , \label{charge}
\end{equation}
where we have chosen the normalization factor to conform to our earlier conventions in TMG. The integral is to be evaluated on a circle at infinity. Let us apply this to the rotating BTZ black hole or a spacetime which is asymptotically a rotating BTZ black hole with the metric
\begin{equation}
ds^2 = (m G_3 + \Lambda r^2 ) dt^2 - a dt d\phi  + r^2 d \phi^2  + \frac{ dr^2}{ - m G_3 - \Lambda r^2 +\frac{a^2}{4 r^2}} 
\end{equation}
where $a$ is the rotation parameter \cite{btz}. Choosing $m=0$ and $a =0$ to be the background, we obtain the mass (energy)  corresponding to the Killing vector $\bar{\xi}^\mu = -(\frac{\partial}{\partial t})^\mu $
 and the angular momentum  corresponding to the Killing vector $\bar{\xi}^\mu = (\frac{\partial}{\partial \phi})^\mu $ of the asymptotically BTZ black hole in MMG respectively as 
\begin{equation}
E =  \frac{1}{G_3} \left ( (\sigma -  \frac{\gamma \Lambda }{2 \mu^2}) m  + \frac{a \Lambda}{ \mu} \right),  \hskip 1 cm J =  \frac{1}{G_3} \left ( (\sigma -  \frac{\gamma \Lambda }{2 \mu^2}) a  - \frac{m}{ \mu} \right).
\end{equation}
Note that these expressions reduce to the TMG forms when $\gamma =0$ \cite{kanik}.  Let us see the chiral gravity limit of MMG from these expressions.  The angular momentum vanishes when 
\begin{equation}
a  = \frac{m}{(\sigma -  \frac{\gamma \Lambda }{2 \mu^2})  \mu},
\end{equation}
and at this point the energy becomes
\begin{equation}
E = \frac{ m}{ G_3 \mu (\sigma -  \frac{\gamma \Lambda }{2 \mu^2}) } M^2_{graviton},
\end{equation}
which also vanishes at the point where the bulk graviton is massless. Note also that positivity of black hole energy is not in conflict with the positivity of energy of excitations.

One might wonder how conserved charges will be defined for spacetimes that are not asymptotically AdS. We shall not go to that discussion here, since it was carried out in \cite{mou,nutku} and the expressions are valid for MMG with small adjustments of the coefficients. 

\section{\label{conc} Conclusions}

Following the Killing charge techniques we have constructed the conserved mass and angular momentum of the recently found Minimal Massive Gravity which only has an on-shell Bianchi Identity.  But that is sufficient to define conserved quantities. We have applied our formulation to the rotating BTZ black hole. We have also studied excitations of the theory and found the left and right central charges of the boundary algebra as well as the energies of massive and massless left and right moving modes. There is no conflict between the boundary and bulk unitarity.  We also constructed the chiral gravity limit of the MMG theory. As of now, it is not clear if this chiral theory has a unitary boundary CFT or not.

This work was supported by  T\"{U}B\.{I}TAK  grant 113F155.

\end{document}